\documentstyle [12pt] {article}

\catcode`@=12
\begin{document}
\vspace{0.5in}
\oddsidemargin -.375in
\newcount\sectionnumber
\sectionnumber=0
\def\be{\begin{equation}}
\def\ee{\end{equation}}
\begin{flushright} UH-511-787-94\\August 1994\\
\end{flushright}
\vspace {.5in}
\begin{center}
{\Large\bf Charm as Probe of New Physics\\}
\vspace{.5in}

{\bf Sandip Pakvasa\\}
\vspace{.1in}
 {\it
Physics Department, University of Hawaii at Manoa, 2505 Correa
Road, Honolulu, HI 96822, USA.}\\
\end{center}


\section{Introduction}

In this talk I would like to discuss two aspects of charm
physics. One is to show that many standard model predictions
for rare decay modes (along with $D^0 -\bar{D}^0$ mixing
and CP violation) are extremely small thus opening a window
for new physics effects\cite{1}; and the other is to review the expectations
from several plausible and interesting new physics
possibilities.

The standard model will be taken to be defined by the gauge
group $SU(3)_c \times SU(2)_L \times U(1)$ with three families of
quarks and leptons, one Higgs doublet and no right handed
neutrinos (thus $m_{\nu_{i}} = 0)$.  We will review
predictions for D mixing, CP violation in the D system and
then discuss rare decays of D's.

Everything in this talk is based upon joint on-going work
with Gustavo Burdman, Eugene Golowich and JoAnne Hewett;
many details and complete results will appear in a
forthcoming review.

\section{$D-\bar{D}$ Mixing and CP Violation}

As already discussed by Burdman,\cite{2} $D^0 \bar{D}^0$ mixing
differs from $K^0-\bar{K}^0$ and $B^0-\bar{B}^0$ mixing in
several ways.  In the box diagram, the s-quark
intermediate state dominates; this is in spite of
the suppression by the factor
$(m_s/m_c)^2$ resulting from the external momenta (i.e. the
fact that $m_c > m_s)$\cite{3}.  The final result for $\delta m$ from
the box diagram is extremely small, one finds

\begin{equation}
\delta m_D \sim \quad 0.5.10^{-17} \ \ GeV
\end{equation}
for  $m_s \sim 0.2$ GeV and $f_D \sqrt{B_D} \sim 0.2$ GeV;
leading to

\begin{equation}
\delta m_D/ \Gamma{_{D^{^0}}} \ \sim \ 3.10^{-5}
\end{equation}

One should worry whether long distance contributions would
give much larger contributions.  The contribution from two
body states $K^+ K^-, \ K^- \pi^+, K^+ \pi^-, \pi^+ \pi^-$ was
carefully evaluated by Donoghue et al. \cite{4}  With the current
experimental values, this is rather small, of the same order
as above.  A very different calculation of the matrix
element resulting from the box diagram due to Georgi et al.
\cite{5}
employing HQET also yields an enhancement of no more than a
factor of 4-5 over the short distance result.  Even if none
of these arguments are completely convincing it is likely
that the SM $\delta m/\Gamma$ is not
enhanced by more than an order of magnitude over the short
distance value of $3.10^{-5}$.  Since the current
experimental limit \cite{6} is 0.083, there is plenty of room for new
physics effects to show up.

CP violation in mixing is described by $\epsilon_D$ and the
asymmetry a in e.g. $e^+e^- \rightarrow D^0 \bar{D}^0
\rightarrow \ell^+ \ell^+ x , \ell^- \ell^- x $ defined by
$a=(N^{++} - N^{--})/ (N^{++}+ N^{--})$
goes as $2Re \ \epsilon_D$ for small $\epsilon_D.  \ 2Re
\ \epsilon_D$ is given by

\begin{equation}
2Re \ \epsilon_D \ = \ \frac{2 Im \ (M_{12} \Gamma_{12}^*)}
{\mid (Im \ M_2) \mid^2 + \mid (Re \Gamma_{12} \mid^2}
\end{equation}

It is always possible to choose a phase convention for the
KM matrix such that $Im \Gamma_{12} = 0$.  Then

\begin{equation}
2 Re \ \epsilon_D \leq \left ( \frac{Im \ M_{12}}{(Re \Gamma_{12})}
\right )
\end{equation}
the left hand side is given by $\left (\frac{m_b m_c}{m_s^2}
\right)^2 \ Im (U_{cb} \ U_{bu^{*}})^2 / \theta_c^2$ and hence

\begin{equation}
2Re \ \epsilon_D \leq \ 10^{-2}. \end{equation}
This is the maximum value for the CP violating charge
asymmetry (due to mixing) in the SM.  The actual value lies
between $5.10^{-3}$ and $5.10^{-4}$.

Direct CP violation can also be looked for in partial rate
asymmetries of charge conjugate states.  Such rate
asymmetries are proportional to $\sin (\phi_i - \phi_j) \sin
(\delta_i - \delta_j)$ where $\phi_i$ are weak CP phases,
$\delta_i$ are final state interaction phases and $i,j$ are
strong interaction eigenstates \cite{7}.  In SM for D (and Ds) decays
there can be no CP violating rate asymmetries for the
Cabibbo allowed decay modes (and for the double
Cabibbo-suppressed modes as well) to the lowest order.  In
Cabibbo-suppressed modes there can be interference between
the quark decay diagram and Penguin (and/or annihilation)
diagram leading to CP violating partial rate asymmetries.
The main difficulty is evaluating the final state
interaction phases.  Several groups have estimated these
phases\cite{8} and based on these the more promising candidates seem
to be $D^+_s \rightarrow K^{*+} \eta (\eta')$ and $D^+
\rightarrow \bar{K}^{*0} K^+ (\rho^0 \pi^+)$ with
asymmetries in the range
of (2-8)$10^{-3}$.

\section{Rare Decays}

There are a number of "rare" (one-loop) decay modes of
D\cite{9}
which have extremely small rates when evaluated in SM; thus
providing a potential window for new physics contributions.

(i)  $D^0 \rightarrow \mu^+ \mu^-$

	At one loop level the decay rate for $D^0
\rightarrow \mu^+ \mu^-$ is given by

\begin{equation}
\Gamma (D^0 \rightarrow \mu^+ \mu^-) \ =
\frac{G_F^4 \ m_W^4 \ f_D^2 \ m_\mu^2 \ m_D \mid F \mid^2}
{32 \pi^3} \sqrt{1-4m^2_\mu / m_D^2}
\end{equation}
where
\begin{eqnarray}
\begin{array}{cl}
F = & U_{us} U_{cs}^* \ (x_s + 3/4 \ x _s^2 \ \ell_n
x _s)
\\
    & U_{ub} U_{cb}^* \ (x _b + 3/4 \ x_b^2 \ell_n
x_b)
\end{array}
\end{eqnarray}
and $x_i = m_i^2/m_W^2$.  This yields a branching
fraction of $10^{-19}$.  There are potentially large long
distance effects; e.g. due to intermediate states such as
$\pi^0, K^0, \bar{K}^0, \eta, \eta')$ or $(\pi \pi, K
\bar{K})$ etc. Inserting the known rates for $P_i
\rightarrow \mu^+ \mu^-$ and ignoring the extrapolation the
result for $B(D^0 \rightarrow \mu^+ \mu^-)$
is $3.10^{-15}$.
This is probably an over-estimate but might give some idea
of the long distance effects.

(ii)  $D^0 \rightarrow \gamma \gamma$

The one loop contribution to $D^0 \rightarrow \gamma \gamma$ can be
calculated in exactly the same way as above and the
amplitude A is found to be approximately $4.6.10^{-14}$
GeV, where $A$ is defined by the matrix element
$A \ q_{1 \mu} \ q_{2\nu} \ \epsilon_{1\rho} \ \epsilon_{2 \sigma}
\ \epsilon^{\mu \nu \rho \sigma}$.

The decay rate is $\Gamma = \mid A \mid ^2m_D^3 / 64 \pi$
and the branching fraction is $10^{-16}.$  The single
particle contributions due to ($\pi, K, \eta, \eta')$
yield $3.10^{-9}$ but again are grossly over estimated.

(iii) $D \rightarrow \nu \bar{\nu} x$.

The decay rate for $c \rightarrow u \nu \bar{\nu}$ (for 3
neutrino flavors) is given by

\begin{equation}
\Gamma = \frac{3 G^2_F \ m_c^5}{192 \pi^3} \left [
\frac{\alpha}{4 \pi x_w} \right ]^2 \mid A_\nu \mid^2.
\end{equation}

Inserting the one loop value for $A_\nu$, one finds for the
branching fractions:
\begin{eqnarray}
\begin{array}{lcl}
B(D^0 \rightarrow \nu \bar{\nu} x) & = &
2.10^{-15} \\
B(D^+ \rightarrow \nu \bar{\nu} x) & = & 4.5.10^{-15}
\end{array}
\end{eqnarray}

For the exclusive modes $D^0 \rightarrow \pi \nu
\bar{\nu}$ and $D^+ \rightarrow \pi^+ \nu \bar{\nu}$ an
estimate of the long distance contributions yields
\begin{eqnarray}
\begin{array}{lcl}
B(D^0 \rightarrow \pi^0 \nu \bar{\nu}) & \sim  & 5.6.10^{-16}
\\
B(D^+ \rightarrow \pi^+ \nu \bar{\nu}) & \sim &
8.10^{-16}
\end{array}
\end{eqnarray}

(iv) D $\rightarrow \bar{K}(K) \nu \bar{\nu}$

These modes have no short distance one loop contributions.
Estimates of long distance contributions.  Estimates of long
distance contributions due to single particle poles yield
branching fractions of the order of $10^{-15}.$

(v) $D \rightarrow \ell \bar{\ell} x.$

The one loop contributions from $\gamma, Z$ and WW
intermediate states give for the inclusive decay mode
$c \rightarrow u \ell \bar{\ell}$ a rate which corresponds to
a branching fraction for $D^+$ of the order
\begin{equation}
B.R. (D^+ \rightarrow \ \ell \bar{\ell} x) \ = 2.10^{-10}
\end{equation}

This corresponds to a fraction for $D^0$ of B.R. $(D^0
\rightarrow \ell \bar{\ell} x) \ = 10^{-10}$.  The exclusive
modes $D^+ \rightarrow \pi^+ \ell \bar{\ell}$ and
$D^0 \rightarrow \pi^0 \ell \bar{\ell}$ are expected to have
somewhat smaller branching fractions in the range of a few
times $10^{-11}.$

(vi) $(D \rightarrow \gamma x.)$

The Penguin diagram can give rise to $c \rightarrow u \gamma$
at one loop level and (before short distance QCD
corrections) gives a rate for $c \rightarrow u \gamma$
corresponding to a branching fraction of B.R.
$(D \rightarrow \gamma x)$ of about $10^{-16}$.
This would yield branching fractions for
exclusive channels such as $D^0 \rightarrow \rho^0 \gamma,
w^0 \gamma$ at a level of $10^{-17}$ or so.  It is expected
that the QCD corrections will enhance this rates (these
calculations are in progress).

On the other hand, if the precise partial wave structure in
the amplitude for the decays such as $D \rightarrow \phi
\rho$ (as well as the total rates) were known, it is possible
to estimate the rates for $D^0 \rightarrow \phi^0 \gamma, D
\rightarrow \rho \gamma$ etc.  At present only upper bounds can be obtained
e.g.
\begin{eqnarray}
\begin{array}{lcc}

B.R. (D^+ \rightarrow \rho^+ \gamma)    & <  &  2.10^{-4}
 \\
B.R. (D^0 \rightarrow \rho^0 \gamma)    & <  &  2.10^{-5}
 \\
B.R. (D^0 \rightarrow \phi \gamma)    & <  &  2.10^{-4}
\end{array}
\end{eqnarray}

If these long distance contributions turn out to be much
larger than the Penguin contributions (even after QCD
correction) then the Penguin will remain invisible in D
decays.  I suspect that this is the case.

 From the data on $D^0 \rightarrow \bar{K}^{*0}
\rho^0$\cite{6} and
VMD one obtains B.R. $(D^0 \rightarrow \bar{K}^{*0} \gamma)
\sim 1.6.10^{-4}$.  From the data on $D^+ \rightarrow
\bar{K}^{*0}
\rho^+$, assuming that $\mid A_1 \mid \gg \mid A_3 \mid$
and that there is no particular enhancement in DCSD mode
$D^+ \rightarrow K^{*+} \rho^0$, one finds B.R.
$(D^+ \rightarrow \bar{K}^{*+} \rho^0) \sim 1.4.10^{-4}$ and
in turn B.R. $(D^+ \rightarrow \bar{K}^{*+} \gamma) \sim
3.10^{-7}$.

I should stress that in all of the above the short distance
QCD corrections have not yet been incorporated.  Since these
tend to enhance the decay rates and the long distance
values tend to be over-estimates, the gap between the two
will be smaller than it appears here.

\section{New Physics Scenarios}

(i)  Additional Scalar Doublet

One of the simplest extensions of the standard model is to
add one scalar Higgs doublet\cite{10}.  If one insists on flavor
conservation there are two possible models:  in one (model
I) all quarks get masses from one Higgs (say $\phi_2)$
and the other $\phi_1$ does not couple to fermions; in the
other $\phi_2$ gives
masses to up-quarks only and $\phi_1,$ to down-quarks only.
The new unknown parameters are $\tan \beta (= v_1/v_2$, the
ratio of the two vevs) and the masses of the additional
Higgs scalars, both charged as well as neutral.

In the charmed particle system, the important effects are in
$\delta m_D$ and the new
contributions due to charged Higgses to rare decays such as
$D^0 \rightarrow \mu^+ \mu^-, D \rightarrow \pi \ell
\bar{\ell}, D \rightarrow \gamma \gamma, D \rightarrow \rho
\gamma$ etc.

The mass of the charged Higgs is constrained to be above
50 GeV by LEP data and there is a joint constraint on $m_H$
and $\tan \beta$ from the observation of $B \rightarrow K^*
\gamma$.  For large $\tan \beta, \ \delta m_D$ can be larger
than the SM results\cite{11}.

(ii)  Fourth Generation

If there is a fourth generation of quarks, accompanied by a
heavy neutrino $(M_{N0} > 50$ GeV to satisfy LEP
constraints) there are many interesting effects observable in
the charm system.

In general $U_{ub'}$ and $U_{cb'}$ will not be zero and then
the $b'$-quark can contribute to $\delta m_D$ as well as to rare
decays such as $D^0 \rightarrow \mu \bar{\mu}, D \rightarrow
\ell \bar{\ell} x, D \rightarrow \pi \nu \bar{\nu}$ etc. (A
singlet b' quark as predicted in E6 GUT has exactly the same
effect).  A heavy fourth generation neutrino $N^0$ with
$U_{eN0} U^*_{\mu N} \neq 0$ engenders decays such as $D^0
\rightarrow \mu \bar{e}$ as well.

For $U_{ub'} U_{cb'} \stackrel{\sim}{>} 0.01$ and $m_{b'}>
100 GeV$, it is found that\cite{12}
\begin{enumerate}
\item[(a)]$\delta m_D/\Gamma \ \ > 0.01$;
\item[(b)] $B(D^0 \rightarrow \mu \bar{\mu}) >
0.5.10^{-11}$;
\item[(c)]B $(D^+ \rightarrow \pi^+ \ell \bar{\ell} \ ) >
10^{-10}$; etc.
\end{enumerate}

For a heavy neutrino of mass $M_{N^0} > 45$ GeV,
the mixing with $e$ and
$\mu$ is bounded by $\mid U_{Ne} U^*_{N \mu} \mid^2 <
7.10^{-6}$\cite{13}
and we find that branching fraction for $D^0 \rightarrow
\mu^- e^+, \mu^+ e^-$ can be no more
than $6.10^{-22}!$  This is also true for a singlet heavy
neutrino unaccompanied by a charged lepton.  To turn this
result around, any observation of $D^0 \rightarrow \mu e$
at a level greater
than this must be due to some other physics, e.g. a
horizontal gauge (or Higgs) boson exchange.

(iii) Flavor Changing Neutral Higgs

It has been an old idea that if one enlarges the Higgs
sector to share some of the large global flavor symmetries
of the gauge sector (which eventually are broken
spontaneously) then it is possible that interesting fermion
mass and mixing pattern can emerge.  It was realized early
that in general this will lead to flavor changing neutral
current couplings to Higgs\cite{14}.  As was
stressed\cite{15} then and has
been emphasized recently\cite{16}, this need not be alarming as long
as current limits are satisfied.  But this means that the
Glashow-Weinberg criterion will not be satisfied and the GIM
mechanism will be imperfect for coupling to scalars.  This
is the price to be paid for a possible "explanation" of
fermion mass/mixing pattern.  Of course, the current
empirical constraints from $\delta m_K, K_L \rightarrow
\mu \mu  \ K_L \rightarrow \mu e$ etc. must be observed.
This is not at all difficult.  For example, in one early
model, flavor was exactly conserved in the strange sector
but not in the charm sector\cite{14}!

In such theories, there will be a neutral scalar, $\phi^0$ of
mass m with coupling such as
\begin{equation}
(g \bar{u} \gamma_5 c \ + \ g' \bar{c} \gamma_5 u)
\phi^0
\end{equation}
giving rise to a contribution to $\delta m_D$

\begin{equation}
\delta m_D \sim \ \frac{g g'}{m^2} \ f^2_D \ B_D \ m_D
\left ( m_D/m_C \right )
\end{equation}
With a reasonable range of parameters, it is easily
conceivable for $\delta m_D$ to be as large as $10^{-13}$
GeV.  There will also new contributions to decays such as
$D^0 \rightarrow \mu \bar{\mu}, D^0 \rightarrow \mu e$ which
will depend on other parameters.

There are other theoretical structures which are
effectively identical to this, e.g. composite technicolor.
The scheme discussed by Carone and Hamilton leads to a
$\delta m_D$ of $4.10^{-15}$ GeV\cite{17}.

(iv) Family Symmetry

The Family symmetry mentioned above can be gauged as well as
global.  In fact, the global symmetry can be a remnant of an
underlying gauged symmetry.  A gauged family symmetry leads
to a number of interesting effects in the charm
sector\cite{18}.

Consider a toy model with only two families and a
$SU(2)_H$ family gauge symmetry acting on LH doublets; with
$\left [(\stackrel{u}{c})_L \ ( \stackrel{d}{s})_L \right ] \  $  and
$\left [(\stackrel{\nu_e}{\nu_\mu})_L   \
(\stackrel{e}{\mu})_L \right ]$
\noindent
assigned to $I_H = 1/2$ doublets.  The gauge interaction
will be of the form:

\begin{equation}
g \left [ (\bar{d} \ \bar{s})_L \ \gamma_\mu \ \bar{\tau}.
\bar{G} \mu \left (
\stackrel{d}{s} \right )_L  \ \ + .......\right ]
\end{equation}
After converting to the mass eigenstate basis for quarks,
leptons as well as the new gauge bosons, we can calculate
contributions to $\delta m_K, \delta m_D$ as well as to
decays such as $K_L \rightarrow e \mu$ and $ D \rightarrow
e\mu$.  The results are:
\begin{equation}
\begin{array}{lcl}
\delta m_D/ \delta m_K & = &
\frac{f_D^2 B_D m_D \left [
\frac{\cos^2 2 \theta u}{m_1^2} \ +
\frac{\sin^2 2 \theta u}{m_2^2} -
\frac{1}{m_3^2} \right ]}
{f_K^2 B_K m_K \left [ d \rightarrow u \right ]}
\\
m(K_L^0 \rightarrow e \mu) & = & \frac{1}{2 \sqrt{2}}
g^2 f_K m_\mu
\left [ \frac{\cos 2 \theta_d cos 2 \theta_e}{m_1^2}
\ +
\frac{\sin 2 \theta_d \sin 2 \theta_e}{m_2^2} \right ]
\bar{\mu} (1 +
\gamma_5) e. \\
m (D^0 \rightarrow e \mu) & = &  \frac{1}{4} g^2f_K m_\mu
\left [ d \rightarrow u  \right ]
\bar{\mu}(1 + \gamma_s) e.
\end{array}
\end{equation}
where $\theta_d, \theta_u$ are $\theta_e$ are the mixing
angles in the $d_L-s_L$, $u_L-c_L$ and $e_L-\mu_L$ sectors
and are not measured experimentally and $m_i$ are the gauge
masses.  It is possible to obtain $\delta m_D \sim 10^{-13}$
GeV and $B (D^0 \rightarrow e \mu) \sim 10^{-13}$ while
satisfying the bounds on $\delta m_K$ and $B(K_L^0
\rightarrow e \mu)$.

(v) Supersymmetry

In the Minimal Supersymmetric Standard Model new
contributions to $\delta m_D$ come from gluino exchange box
diagram and depend on squark mixings and mass splittings.
To keep $\delta m_K^{SUSY}$ small the traditional ansatz has
been squark degeneracy.  In this case $\delta m_D^{SUSY}$ is
also automatically suppressed, no more than $10^{-18}$ GeV
\cite{19}.  Recently it has been proposed\cite{20} that
another possible way to keep $\delta m_K^{SUSY}$ small is to
assume not squark degeneracy but proportionality of the squark
mass matrix to the quark mass matrix to the quark mass
matrix. It turns out in
this case that $\delta m_D$ can be as large as the current
experimental limit.  In some non-minimal SUSY theories
certain radiative decay modes can have large rates\cite{21}.

(vi)  Left-Right Symmetric Models

In a very nice paper\cite{22}, the Orsay group has pointed
out that in left-right symmetric extensions of the SM, there
can be sizable CP violating asymmetries in the Cabibbo
allowed decay modes (which is impossible in the SM).  I
would like to illustrate this but in a different kind of
model, the model of Gronau and Wakaizuni\cite{23}.

Recall that the basic premise of the model is that the
suppression of $b \rightarrow c \ell \nu$ decays is not due
to a small mixing $U_{bc}$ but due to the decay proceeding
via $W_R$ exchange and the smallness of the ratio
$(m{_{W_{L}}}/m{_{W_{R}}})^2$.
This is accomplished by enlarging
the gauge group to
$SU(2)_L \times SU(2)_R \times U(1)$ but without manifest left-right
symmetry and assuming the two mixing matrices to be

\begin{equation}
U_L \ = \ \left (
\begin{array}{ccc}
1               & \lambda            & \rho \lambda^3 \\
^-\lambda       &  1                 &   0    \\
-\rho \lambda^3 &  - \rho \lambda^4  &  1
\end{array}   \right )
\end{equation}

\begin{equation}
U_R = \left (
\begin{array}{ccc}
e^{i \alpha}     &  0           &  0 \\
0                & s e^{i\delta}     & c e^{i \beta} \\
0                & c e^{i \gamma} &  - s e^{i(\beta \ +
\gamma)}
\end{array}  \right )
\end{equation}
where $\lambda
$ and $\rho$ are the usual Wolfenstein
parameters and $U_L$ is real.  As is evident, the current $b
\rightarrow c$ is pure RHC.  For successful phenomenology
and a good fit to all the data there are a number of
constraints on the model; e.g. $\nu_R$ must have a mass in
the range of few MeV, $\rho \sim 0.2$ to 0.7, $m{_W{_R}} > 400$
GeV, $c > 0.8, s < 0.6$.  All CP violation comes from the RH
sector and $\epsilon$ and $\epsilon'$ require that:  $\sin
(\gamma - \alpha) > 0.1, \sin (\delta - \alpha) < 0.5$
and $\sin(\alpha + w) < 0.7$; thus the constraints on the
phases in $U_R$ are rather weak.

In this model, for a decay such as $D \rightarrow \bar{K}
\pi$, in addition to the $W_L$ mediated decay there is an
additional amplitude due to $W_R$ which now carries a CP
phase.  Because of the larger $W_R$ mass, the QCD
coefficients for the $RR$ operators are different from the
$LL$ operators resulting in a different ratio for the $I=
3/2$ to $1/2$ final states from the two operators; hence
$a_{3R}/a_{1R} \neq a_{3L}/{a_{1L}}$.  Then the CP partial rate
asymmetry for the decay mode $D^0 \rightarrow K^- \pi^+$ and
$\bar{D}^0 \rightarrow K^+ \pi^-$ is given by

\begin{equation}
\frac {\Gamma-\bar{\Gamma}}{\Gamma + \bar{\Gamma}} =
\frac {s (a_{3R} - a_{1R})}{a_{1L}} \  \sin
(\delta_1-\delta_3) \ \sin (\alpha -\delta)
\end{equation}
where we have taken from data $a_{1L} \sim a_{3L}$.  If, for
simplicity, we take $a_{1R} \gg a_{3R}$, then the RHS
becomes
\begin{equation}
s \left (m{_{W_L}}/m{_{W_R}}\right )^2  \ \ \sin
(\delta_1 - \delta_3) \sin (\alpha - \delta).
\end{equation}

Taking $s \sim 0.5$, $\sin (\alpha - \delta) \sim 0.5$ in the
allowed range, $\delta_1 - \delta_3 \sim 0(90^0)$ from data,
and $(m{_{W_{L}}}/m{_{W_{R}}})^2 \sim 0.04$ the
asymmetry is of the order 0.01 to be compared to 0 in SM.
As shown in Ref. \cite{22} similar values obtain in other
left-right symmetric models as well making this a generic
result in Left-Right Symmetric theories.  Incidentally, the
new contributions to $\delta m_D$ are no larger than in SM.

\section{\bf Conclusion}

To summarize, in the charm system several phenomena (such as
$\delta m$, CP, loop induced decays) which are easily
observed in K and B system are greatly suppressed in SM and
there is a window of opportunity for new physics to show up.

Of course, even when there is new physics beyond the
standard model (BSM) it is not guaranteed that there are
interesting signals large enough to be seen.  Probably the
most likely place for some new physics to show up in $\delta
m_D$.  To disentangle the origin some other effects have to
be seen.  CP violation (in channels forbidden in SM) and
rare decays such as $D^0 \rightarrow \mu \bar{\mu}, \gamma
\gamma, \nu \nu x$ etc.  would come a close second.  Decays
such as $D^0 \rightarrow \mu e$ are probably unlikely to
occur at rates large enough to be seen in the near future
but who knows?

\section{\bf Acknowledgements}

I am most grateful to my collaborators Gustavo Burdman,
Eugene Golowich and JoAnne Hewett for providing insight and
wisdom.  I thank Dan Kaplan for the invitation to
participate in CHARM2000 and the Rare Decay, CP and $\delta
M$ Working Groups for enjoyable and stimulating discussions.
 This work was supported in part by USDOE Grant \#DE-FG
03-94ER40833.


\begin{thebibliography}{99}

\bibitem {1}
I. Bigi, Charm Physics Symposium, Beijing, 1987, p. 339;
K. S. Babu, X-G He, X-Q. Li and S. Pakvasa, Phys. Lett. {\bf
B205}, 540 (1988);
X-H. Gao and X-Q. Li,  BIHEP-TH-90, S. Egli {\it et al.,}
ETZ-IMP PR/92-1;
A. Le Yaouanc {\it et al.,} LPTHE Orsay 92/49; A.J. Schwartz,
Mod. Phys. Lett. {\bf A8}, 967 (1993).

\bibitem {2}
G. Burdman, These Proceedings.

\bibitem {3}
A. Datta and D. Kumbhakar, Z. Phys. {\bf C27} (515) 1985; H.
-Y. Cheng, Phys. Rev. {\bf D26}, 143 (1982).

\bibitem {4}
J. Donoghue {\it et al.,} Phys. Rev. {\bf D33}, 179 (1986).

\bibitem {5} H. Georgi, Phys, Lett. {\bf 297}, 353 (1992);
T. Ohl {\it et al.,} Nucl. Phys. {\bf B403}, 605 (1993).

\bibitem {6} K. Hikasa {\it et al.,} PDG, Phys.
Rev. {\bf D45}, S1 (1992).

\bibitem {7}T. Brown {\it et al.,} Phys. Rev. Lett. {\bf
51}, 1823 (1983).

\bibitem {8} L-L. Chau and H-Y. Cheng, Phys. Rev. Lett. {\bf
53}, 1037 (1984); F. Buccella {\it et al.,} Phys. Lett. {\bf
B302}, 319 (1993).

\bibitem {9} The basic formulae for these are given in T.
Inami and C. S. Lim, Prog. Theoret. Phys. {\bf 55}, 297
(1981).

\bibitem {10} L. F. Abbott {\it et al.,} Phys. Rev. {\bf
D21}, 179 (1980); V. Barger {\it et al.,} Phys. Rev. {\bf D41},
3421 (1990).

\bibitem {11} J. Hewett (Private Communication).

\bibitem {12} K. S. Babu {\it et al.,} Phys. Lett. {\bf
205B}, 540 (1988).

\bibitem {13} A. Acker and S. Pakvasa, Mod. Phys. Lett. {\bf
A7}, 1219 (1992).

\bibitem {14} S. Pakvasa and H. Sugawara, Phys. Lett.
{\bf B73}, 61 (1978).

\bibitem {15} S. Pakvasa, H. Sugawara and Y. Yamanaka, Phys.
Rev. {\bf D25}, 1895 (1982).

\bibitem {16} L. Hall and S. Weinberg, Phys. Rev. {\bf D48},
979 (1993); S. Pakvasa, Discovery of Neutral Currents, Santa
Monica, CA  1993; AIP Conf. Proc. 300, Ed. A. K. Mann and D.
Cline AIP (1994), p. 426.

\bibitem {17} C. D. Carone and R. T. Hamilton, Phys. Lett.
{\bf 301}, 196 (1993).

\bibitem {18} G. Volkov {\it et al.,} Yad. Fiz. {\bf 34}, 435
(1981).

\bibitem {19} J. S. Hagelin {\it et al.,} Nucl. Phys. {\bf
B415}, 293 (1994) and references therein.

\bibitem {20} Y. Nir and N. Seiberg, Phys. Lett. {\bf B309},
337 (1993).

\bibitem {21} I. Bigi {\it et al.,} Z. Phys. {\bf C48}, 633
(1990).

\bibitem {22} A. Le Yaouanc {\it et al.,} Phys. Lett. {\bf
B292}, 353 (1992).

\bibitem {23} M. Gronau and S. Wakaizumi, Phys. Rev. Lett.;
{\bf 68}, 1814 (1992); See also W-S. Hou and D. Wyler, Phys.
Lett. {\bf B292}, 364 (1992); T. Hattori et al,
TOKUSHIMA-93-06.

\end{thebibliography}
\end{document}